\documentstyle[12pt]{article}
\newcommand{\beq}{\begin{equation}}
\newcommand{\eeq}{\end{equation}}

\def\half{{\textstyle{1\over2}}}

\def\p1half{{\textstyle{{{p+1}\over{2}}}}}

\def\23phalf{{\textstyle{{{23-p}\over{2}}}}}

    \let\p=\pi

 \def\bd{\begin{document}} \def\ed{\end{document}}
\def\ds{\documentstyle} \let\fr=\frac \let\bl=\bigl \let\br=\bigr
\let\Br=\Bigr \let\Bl=\Bigl
\let\bm=\bibitem
\let\na=\nabla
\let\pa=\partial \let\ov=\overline
\newcommand{\be}{\begin{equation}}
\newcommand{\ee}{\end{equation}}
\def\ba{\begin{array}}
\def\ea{\end{array}}
\def\ft#1#2{{\textstyle{{\scriptstyle #1}\over {\scriptstyle #2}}}}
\def\fft#1#2{{#1 \over #2}}
\def\del{\partial}
\def\sst#1{{\scriptscriptstyle #1}}
\def\oneone{\rlap 1\mkern4mu{\rm l}}
\def\ie{{\it i.e.\ }}
\begin{document}
\thispagestyle{empty}
\begin{titlepage}

\bigskip
\hskip 3.7in{\vbox{\baselineskip12pt
}}

\bigskip\bigskip
\centerline{\large\bf CHL Compactifications Revisited}

\bigskip\bigskip
\bigskip\bigskip
\centerline{\bf Shyamoli Chaudhuri\footnote{Email: shyamolic@yahoo.com}}
\bigskip
\centerline{1312 Oak Drive}
\centerline{Blacksburg, VA 24060}
\bigskip\bigskip

\bigskip
\begin{abstract}
CHL compactifications are supersymmetry preserving orbifolds of any
perturbatively renormalizable and ultraviolet finite ground state of the perturbative 
string theories: heterotic, type 
I, or type II, preserving 32, 16, 12, 8, 4, (or zero) supersymmetries, and retaining
the perturbative renormalizability and finiteness of the parent string vacuum.
In this paper, we review the genesis of the 
CHL (Chaudhuri-Hockney-Lykken) project within the broader context
of the full String/M Duality web, establishing the existence of 
moduli spaces with a {\em small} number of 
massless scalar fields, the 
decompactification of such moduli spaces to one of the five 
ten-dimensional superstring theories, and the
appearance of electric-magnetic duality in only the {\em four-dimensional}
moduli spaces, a 1995 observation due to Chaudhuri \& Polchinski.
We present two mathematical curiosities easily deduced from the
fermionic current algebra representation but whose physical significance
is a puzzle: a 4D N=4 heterotic 
string vacuum with no massless scalar fields other than the dilaton, and
a 2D N=8 heterotic string vacuum with no abelian gauge fields, reiterating 
once more the necessity for a systematic classification of the CHL orbifolds.
\end{abstract}

\end{titlepage}

\section{Introduction}

\vskip 0.1in An important direction of current research in String
Theory is to determine
its precise boundaries, thereby discovering the 
principles by which we can eliminate the redundancy of the 
multiplicity of string vacua and arrive at a convincing description of the
real world. This has been the theme of my research for many years, and
it grew out of the 1995 discovery of the CHL strings
\cite{cp,chl}. This paper offers a retrospective on this discovery within the 
larger context of the String/M Duality web, highlighting some 
of the results and follow-up insights that I have found 
especially significant in succeeding years.\footnote{Two recent workshop presentations 
by me which
overlap with some parts of this paper can be found in \cite{land,chltalk}.}

\vskip 0.1in The CHL strings are supersymmetry 
preserving orbifolds of any consistent compactification of
perturbative superstring theory, where by consistency we mean here
an exactly solvable background of perturbative string theory 
to all orders in the $\alpha^{\prime}$ expansion. Such backgrounds 
have a solvable superconformal field theory description on the worldsheet, 
leading to an anomaly-free, and ultraviolet finite, perturbatively 
renormalizable
superstring theory in target spacetime. We use the term {\em perturbatively 
renormalizable}  to describe such a target spacetime string theory Lagrangian 
despite the presence of  infinitely many couplings in the $\alpha^{\prime}$ expansion, 
because {\em only a finite number of independent parameters} go into their determination, 
and these can all be found at the lowest orders in the string effective Lagrangian.
The existence of only a finite number of independently renormalized couplings
is the defining criterion for the Wilsonian renormalizability of a quantum theory.
Thus, from this 
perspective, the perturbative string theoretic unification of gravity and Yang-Mills gauge
theories with chiral matter can be seen as providing a precise, and unique, 
gravitational extension of the 
anomaly-free and renormalizable Standard Model of Particle Physics. 

\vskip 0.1in Of course, since we lack a precise formulation of 
nonperturbative string theory at the
current time, we can only reliably invoke the above 
framework as long as we remain within the domain of weak string coupling.
Fortunately, all observational signals point to the weak unification of the gauge
couplings and gravity in our four-dimensional world, with the supersymmetry breaking 
scale lying somewhere between the electroweak (TeV) and gauge coupling unification 
($10^{16-17}{\rm GeV})$ scales. So perhaps we will be lucky and able to
follow the target spacetime string effective Lagrangian approach up to at 
least the gauge coupling unification scale, using tried-and-true renormalization group methods.
Indeed the stream of precision data from the Z factories in the early 90's, pinning down 
both the number of lepton-quark generations, as well as the
hierarchical texture of fermion masses with the discovery of a surprisingly heavy top quark, 
and increasingly tight windows on neutrino masses, stimulated the resurgence of
theoretical investigations of supersymmetric grand unification models using such
renormalization group techniques. 

\vskip 0.1in It was in light of
these developments that Joe Lykken and I initiated an ambitious new effort in string model
building in 1993-94. Our original goal was to identify
exactly solvable conformal field theory (cft) realizations of four-dimensional heterotic string vacua 
with massless particle spectra and couplings that would cover the spread of plausible
semi-realistic extensions of the supersymmetric Standard Model, perhaps even suggest
some new features unforseen in conventional, field theoretic model building. We focussed
on fermionic current algebra realizations of the conformal field theory, using a formalism originally
developed
by Kawai, Lewellen, Schwartz, and Tye \cite{klst}, because of the 
simple, and explicit, nature of this description. It is straightforward to {\em embed} the
desired particle spectrum and couplings in the cft using the fermionic representation, 
the symmetries of the desired low energy string 
spacetime effective Lagrangian become transparent. Each of these exact cft solutions correspond to 
special points in the moduli space of some CHL orbifold with a chiral, 4D N=1 supersymmetry. 
The detailed checks of the worldsheet constraints in such N=1 string
vacua are
prohibitively calculation-intensive.
The algorithms were therefore implemented in an interactive and user-friendly
computer program, {\em Spectrum}, created largely by George Hockney at 
Fermilab,
and aimed at facilitating phenomenological string model building.
Remnant ambiguity in the implementation of 
modular invariance for the fermionic twisted $Z_2$ current algebras was resolved by
invoking the Verlinde fusion rules, an analysis due to Joe
Lykken, in collaboration with a Fermilab postdoc, Stephen-wei Chung \cite{consist}.

\vskip 0.1in
Let me outline the
phenomenological successes of just one of our 4D N=1 examples, designated 
the CHL5 Model in the literature \cite{chl2,cvet}. It describes an N=1 heterotic
string vacuum
with three generations of supersymmetric Standard Model 
$SU(3)_C$$\times$$SU(2)_L$$\times$$U(1)_Y$ particles, an anomalous
$U(1)$, and a very small number of flat directions at the string scale. Analysis
of the flat directions removes all but one additional $U(1)$ at the string scale
in the anomaly-free vacuum, the hypercharge embedding mimics the $SU(5)$ 
result extremely well, $k_Y$$=$$11/6$, quite close to $5/3$, without actual 
grand unification, giving acceptable values for the gauge coupling unification
scale. One generation is singled out from the other two by its distinct couplings
already at the string scale. The breaking of the additional $U(1)$ at either an 
intermediate, or electroweak, scale can generate interesting fermion mass 
textures. The detailed implications of either scenario for CHL5 
have been explored in the papers by Cleaver, Cvetic, Espinosa, Everett, 
and Langacker \cite{cvet}. 

\vskip 0.1in It should be emphasized that CHL5 is already a rather good
string theory description of the observable sector of supersymmetric standard model 
particle physics. But the hidden sector of this model is rather heavily constrained, and
not terribly interesting. Helpful
new input that would enable one to improve on such model-building exercises 
is expected to come in the near future when LHC turns on, giving
insight into both the supersymmetry breaking scale as well as, hopefully, the mass of the 
lightest supersymmetric partner. Inputs from the crucial neutrino sector are already
the focus of
both current, and future, astro-particle experiments. Perhaps whole 
classes of inflationary models can be ruled out so
that we will have additional insight both into viable mechanisms for supersymmetry
breaking, as well as the viable early Universe scenarios.
I should emphasize that all of this is nothing more than physics
extracted from the {\em leading} terms in the string 
spacetime effective Lagrangian: the infrared limit of the full perturbative
string theory, and in a specific 4D flat target spacetime background. 

\vskip 0.1in In the 
intervening years since 1995, investigations of string phenomenology 
have focussed on
the incorporation of warped target spacetime metrics, background 
fields and 
fluxes, including supergravity pform fluxes from 
both Neveu-Schwarz and Ramond sectors, and their 
consequences for the hierarchy problems of particle physics,
for moduli stabilization, 
supersymmetry breaking, and the form of the inflationary
potential. We should note that, with some exceptions, these theoretical developments 
have largely 
been restricted to analysis of the low energy string spacetime effective Lagrangian, 
and the implications of $\alpha^{\prime}$ corrections have not always been clarified.
But it should be evident that any generic insights from such analyses will
apply equally well 
to the phenomenology of the CHL orbifolds. We are omitting detailed references for 
brevity, deferring discussion to future work.

\vskip 0.1in 
The nonperturbative, pre-spacetime-geometry framework for string/M theory that 
may lie below the 
distance scale at which target spacetime Lagrangians become
our primary investigational tool is addressed in my recent papers \cite{mtheory,land}. 
A rather important question that needs to be addressed at this
juncture 
is the apparent {\em disconnectedness} of the vacuum landscape of String Theory
evidenced by the discovery of the CHL orbifolds \cite{chl,cp}, a
phenomenon that raises the spectre of both island Universes, and of a
fundamental role for the 
Anthropic Principle \cite{ap}. 
The theoretical paradigm that will enable us to understand these issues 
is provided by the Hartle-Hawking framework for 
Quantum Cosmology \cite{hawk}, but let me begin by explaining the
nature of the moduli spaces of the CHL compactifications. 

\section{A Brief Introduction to the CHL Strings}

\vskip 0.1in The CHL (Chaudhuri-Hockney-Lykken) strings were
named by Polchinski \cite{cp} for the 
authors of the 1995 paper that pin-pointed the existence of additional exactly solvable
 supersymmetry
preserving solutions to the heterotic string theory consistency 
conditions other than toroidal compactifications \cite{chl}.
As explained above, the original motivation for the study of the CHL compactifications was better 
4D N=1 low energy susy particle phenomenology, and a serious wrinkle on such
efforts had been the proliferation of massless scalar moduli in any semi-realistic 
examples. To understand why the CHL compactifications have many fewer flat
directions, consider the simplest example, the 
 Chaudhuri-Polchinski 
orbifold of the circle compactified 
$E_8$$\times$$E_8$ heterotic string described in \cite{cp}:
\begin{equation}
p = {{1}\over{{\sqrt{2}} }} (p_1 + p_2 )  \in \Gamma_8 , \quad  
p = {{1}\over{{\sqrt{2}} }} (p_1 - p_2 )  \in \Gamma_8^{\prime} , \quad p_3 \in \Gamma^{(1,1)} 
\quad ,
\label{eq:mom}
\end{equation}
Let us mod out by the ${\rm Z}_2$ outer automorphism, $R$, interchanging
the two $E_8$ lattices, $\Gamma_8$$\oplus$$\Gamma_8^{\prime}$,
accompanied by a translation, $T$, in the (17,1)-dimensional momentum lattice,
$({\bf v}, 0 ; {\bf v}_3)$ \cite{cp}. This projects onto the symmetric linear 
combination of the momenta in the two $E_8$ lattices, so that the gauge
group is generically $E_8$$\times$$U(1)$:
\begin{equation}
p = {{1}\over{{\sqrt{2}} }} (p_1 + p_2 )  \in \Gamma_8 , \quad  
p = {{1}\over{{\sqrt{2}} }} (p_1 - p_2 )  \in \Gamma_8^{\prime} , \quad p_3 \in \Gamma^{(1,1)} 
\quad ,
\label{eq:mome}
\end{equation}
The RT orbifold acts on the perturbative heterotic string spectrum 
as follows \cite{cp}. Let ${\cal U}$ and ${\cal T}$ denote, respectively,
the untwisted and twisted sectors of the orbifold. The untwisted sector is
composed of states invariant under R (${\cal I}$), and under T (${\cal I}^*$):
\begin{eqnarray}
{\cal I}:&& \quad \quad p_1 = 0 , \quad p_2 = {\sqrt{2}} \Gamma_8 , \quad 
p_3 \in \Gamma^{(10-d,10-d)} 
\cr
{\cal I}^*: &&\quad \quad p_1 = 0 , \quad p_2 = {{1}\over{{\sqrt{2}} }} 
\Gamma_8 , \quad 
p_3 \in \Gamma^{(10-d,10-d)} \cr
{\cal T}: && \quad \quad 
{\bf p} \in {\cal I}^* + {\bf v} \quad . 
\label{eq:orb}
\end{eqnarray}
Notice that the dimension of the moduli space is much {\em smaller}:
the massless scalars parametrize the coset, $SO(18-d,10-d)/SO(18-d)\times SO(10-d)$,
upto discrete identifications, for compactification on the torus $T^{10-d}$. 
The momentum
vectors in the Hilbert space of the orbifold lie on hyperplanes within the 
(26-d,10-d)-dimensional lattice describing the toroidally compactified
$E_8$$\times$$E_8$ string. Such moduli spaces include several novelties
including affine Lie algebra realizations of the simply-laced gauge groups
at higher Kac-Moody level, as well as enhanced symmetry points with non
simply laced gauge symmetry \cite{chl,cp}. We will return to these novelties below.

\vskip 0.1in 
Notice that since the orbifold
action in question {\em preserves} supersymmetry, our discussion of the disconnectedness of
the CHL moduli spaces with 16 supersymmetries can be carried over to 
an analogous disconnectedness of CHL moduli spaces constructed 
as orbifolds of 4D heterotic vacua with 12, 8, 4, or even zero, supersymmetries. 
Consider  the decompactification limit to ten dimensions 
without any change in the number of supersymmetries: for the $Z_2$ orbifold, 
the outer automorphism interchanging the two $E_8$ lattices
becomes trivial in this limit, and we straightforwardly recover an additional $248$
massless gauge bosons. A subsequent toroidal compactification completes the 
continuous interpolating path connecting a point in the moduli space of a
CHL orbifold with some point
in the moduli space of toroidal compactifications, via the ten-dimensional 
$E_8$$\times$$E_8$ heterotic string ground state. Thus, a spontaneous decompactification to
ten dimensions followed 
by a spontaneous re-compactification can indeed interpolate
between a pair of (dis)connected CHL orbifolds with different numbers of abelian 
multiplets. But this is a genuinely stringy phenomenon, with a slew of
modes with string scale masses descending into the massless field theory as we tune
the compactification radius to its noncompact limit.

\vskip 0.1in A similar example is the spontaneous restoration of extended supersymmetry
known to occur in certain decompactification limits of the moduli space of 
4D $N$ $=$ $2$ heterotic string compactifications \cite{koun1}. The modular invariant one
loop vacuum amplitude of the freely acting orbifold in question is parameterized by
the continuously varying, complex structure moduli and Kahler moduli of a six-torus, in addition 
to a constant background electromagnetic field; the
extended supersymmetry is restored in the limit that one of the cycles of the torus
decompactifies \cite{koun1}. Does this framework allow for continuous
interpolations between the moduli space of toroidal compactifications of the 
heterotic string and a CHL compactification with eight fewer abelian gauge fields along
a path that traverses a family of ground states with only eight supercharges, 
and in one lower spacetime dimension? 

\vskip 0.1in The problem with either of these proposals is that the interpolating 
trajectories are exactly marginal flows from the perspective of the 2D worldsheet
renormalization group. Thus, there is no reason to expect the stringy ground state to
\lq\lq evolve" along such a trajectory in the absence of supersymmetry breaking with
a consequent lifting of the vacuum degeneracy. In other words, if the 
supersymmetry breaking scale in Nature
does turn out to be significantly lower than the string scale, the stringy massive 
modes in the CHL orbifold will have genuinely decoupled from the low energy 
field theory limit, and there is no escaping the conclusion that the field theoretic dynamics 
of vacuum selection occurs in one of a multitude of disconnected, low energy Universes. 
How should we
interpret the resulting multitude of low energy string 
effective Lagrangians? The Hawking-Hartle paradigm \cite{hawk} would
identify each such low-energy spacetime 
effective Lagrangian as the final state of a {\em consistent 
history} in some putative Quantum Theory of the Universe. The pre-spacetime
matrix framework for nonperturbative String/M theory described in my recent work 
\cite{land} is such a theory, yielding also a multitude of acceptable 
spacetime effective Lagrangians, each characterized by a distinct large N limit 
of the matrix
Lagrangian. The \lq\lq theory" for the
Initial Conditions of the Universe \cite{hawk}, to borrow a phrase from 
Hartle and Hawking, 
is the pre-spacetime finite N matrix dynamics. This dynamics is 
beyond the direct purview of perturbative string theory.

\vskip 0.1in To summarize, if the supersymmetry-breaking scale 
is clearly separated from the string mass-scale, the stringy massive modes will 
have genuinely decoupled from the effective Lagrangian of relevance and there 
is no escaping the conclusion that vacuum 
selection in perturbative string
theory involves more than just 
dynamics, requiring a discrete choice among disconnected low energy Universes. 
However, upon including the 
stringy massive modes, 
all of the CHL orbifolds are connected in the sense that they
decompactify to the same 10d perturbative string vacuum. 

\vskip 0.1in
By now, the CHL strings have given many fundamental new insights into 
weak-strong electric-magnetic dualities in the String/M theory web \cite{chl,flux,land}.
Let me mention the earliest of these discoveries which appears in the paper
\cite{cp}; this particular observation is due to Joe Polchinski. Careful examination of
which non-simply laced gauge groups can appear at the enhanced symmetry 
points in the moduli space of the CHL orbifold reveals the result \cite{cp}:
\begin{equation}
{\rm Sp} (20-2n) \times SO(17-2d+2n) \quad n=0, \cdots , 10-d 
\quad ,
\label{eq:esp}
\end{equation}
at special points within the same d-dimensional moduli space. Remarkably, the
electric and magnetic dual groups, ${\rm Sp}(2k)$ and $SO(2k+1)$ for given $k$, 
only appear together in the moduli spaces of the four-dimensional CHL orbifolds \cite{cp}. This is 
precisely as required by the S-duality of the 4D N=4 theories, constituting 
independent evidence in favor of it. It should be noted that this property follows
as a consequence of the constraints from modular invariance on the orbifold 
spectrum, the worldsheet constraints responsible for the perturbative renormalizability and 
ultraviolet finiteness of the CHL compactifications. As mentioned above, to the 
best of my knowledge, all of the CHL orbifolds 
described in \cite{cp,cl,land} decompactify to one of the five 10d superstring
theories. A classification of the supersymmetry preserving automorphisms of 
Lorentzian self-dual lattices up to lattices of dimension (22,6) would completely pin down
this important issue, also enabling a classification of the enhanced symmetry
points in each moduli space. This is crucial information necessary for any further
exploration of electric-magnetic duality in the 4D CHL orbifolds. 

\vskip 0.1in
My work on the abelian symplectic orbifolds of six- and four-dimensional 
toroidally compactified heterotic strings  
with David Lowe in \cite{cl} utilized Nikulin's classification of the supersymmetry 
preserving automorphisms of
(19,3)-dimensional Lorentzian self-dual lattices, namely, the cohomology lattices of the 
classical K3 surfaces. Our analysis proceeds as follows: begin at a point in the
moduli space where the (22,6)-dimensional heterotic momentum lattice decomposes  
as $\Gamma^{(19,3)}$$\oplus$$\Gamma^{(3,3)}$. Given Niemeier's enumerative list of 
self-dual lattices up to dimension 24, one can straightforwardly enumerate a large 
number of CHL orbifolds by invoking Nikulin's classification \cite{cl}. For
instance, the ${\rm Z}_2$ orbifold 
described above readily generalizes to ${\rm Z}_n$ orbifolds with $n$$>$$2$
whenever the (19,3) lattice contains $n$ identical component root-lattices. Modding by
the ${\rm Z}_n$ symmetry under permutation, accompanied by an order-$n$ shift vector 
in the (3,3) torus, gives a ${\rm Z}_n$ CHL orbifold. It is evident that 
a classification of the 4D CHL orbifolds would require 
extending Nikulin's analysis to a classification of the symplectic automorphisms of 
all (22,6)-dimensional Lorentzian self-dual lattices. 

\section{Moduli Spaces of Heterotic--Type I--Type II CHL Strings}

\vskip 0.1in
The detailed picture of the
string {\em Landscape} with sixteen supercharges given by the study 
of the CHL compactifications 
can go a long way towards determining the precise boundaries of the
String/M Duality web. We will now explain how this systematic approach can be
successfully applied both to a study of the string 
landscape with 12, 8, 4, or 0 supercharges, as well as to any of the string
theories, heterotic, type I, or type II \cite{flux,land}. 

\vskip 0.1in The heterotic and type IB string theories with gauge group $SO(32)$ are
related by a strong-weak coupling duality transformation in ten dimensions \cite{polwit}. 
The strong coupling limit of the 10d $E_8$$\times$$E_8$ heterotic string is, instead,
conjectured
to be {\em eleven}-dimensional M theory compactified on $S^1/{\rm Z}_2$ \cite{witdual,hw}.
We will begin this section by explaining the nature of the moduli spaces of perturbatively
renormalizable string ground states with sixteen supersymmetries, obtained by either 
toroidal and supersymmetry-preserving orbifold (CHL) compactifications \cite{nsw,chl,cp,cl,flux}, or
asymmetric orbifold and K3 compactification \cite{koun,k3,cl} of, respectively, 
heterotic and type I, or
type IIA and type IIB, superstring theories. The generic moduli space in either case is of 
CHL type \cite{cp,cl}. 

\vskip 0.1in
It is helpful to begin by considering the target spacetime and strong-weak coupling
dualities that relate
the circle-compactified 
type I and heterotic string theories in {\em nine} dimensions and below. The reason
is that, in nine dimensions, 
the two heterotic string 
theories share a common moduli space and one can smoothly interpolate between
ground states with enhanced gauge symmetry $SO(32)$, $SO(16)$$\times$$SO(16)$,
and $E_8$$\times$$E_8$,
respectively, by turning 
on an appropriate Wilson line wrapping the spatial coordinate $X^9$ \cite{nsw,ginine}:
the two heterotic string theories are related by a T-duality transformation on $X^9$.
Likewise, the type IB string theory with 32 D9branes and $SO(32)$ gauge fields
can be mapped by a T-duality on $X^9$ to the type I$^{\prime}$
string theory with 32 D8branes and identical gauge group.  The strong coupling limit of the
latter theory is M theory compactified on $S^1$$\times$$S^1/{\rm Z}_2$, with nonabelian
gauge fields on the domain walls bounding the interval. It should be noted that this 
strong-weak duality has been conjectured for type I$^{\prime}$ ground states with either 
gauge group $SO(16)$$\times$$SO(16)$, or with the extension to a full 
$E_8$$\times$$E_8$. 

\vskip 0.1in The latter enhanced symmetry point corresponds to a 
{\em nonperturbative} 9D background of
the type I$^{\prime}$ 
string theory: one must introduce a pair of D0branes, in addition to the 16 D8branes on each of two
orientifold $O8$-planes.
This brane-configuration preserves all of the supersymmetries of the type I$^{\prime}$ 
string \cite{bachas,polbook,flux}.
The crucial massless gauge bosons in the 
spinor representation of $SO(16)$$\times$$SO(16)$ necessary for the enhancement to
$E_8$$\times$$E_8$ appear as follows. Consider the step-wise change in the background
value of the Ramond-Ramond zero form field strength, $F_0$, associated with the 
creation of a fundamental string \cite{bachas} when a D0brane, and image, threads the
stack of 8 D8branes, plus image-branes, at either orientifold $O8$-plane. The change 
in Ramond
Ramond zero-form flux at each crossing can take either sign $\pm$,
and the absence of dilaton gradients requires that the {\em net} change including 
that at both
orientifold-planes, is zero.
It is easy to verify that the {\bf 128} states in the spinor
of $O(16)$ correspond to the distinct sequences of changes in
the zero form flux at either $O8$-plane that can satisfy this condition.
Notice the isomorphism with
the standard parameterization of the $E_8$ momentum lattice in the 
heterotic string \cite{polbook}: denote the sequence of changes in zero form flux
at an $O8$-plane by a vector as follows, $(\pm , \pm , \pm , \pm , 
\pm , \pm , \pm , \pm )$, including all sequences with an
 even number of minus signs, $\bf 2$$+$$\bf 56$ $+$$ \bf 70$,
and with the compensating sequence at the other $O8$-plane \cite{flux}. 
As was emphasized by us in \cite{flux}, since
the $SO(32)$ and $E_8$$\times$$E_8$ heterotic string ground states are 
continuously connected in nine dimensions and below, self-consistency with the
strong-weak coupling duality relation linking heterotic and type IB $SO(32)$ strings in 
ten dimensions, {\em requires} an enhanced symmetry point in nine dimensions
with $E_8$$\times$$E_8$ gauge fields in the moduli space of the circle compactified
type I-I$^{\prime}$ string theory.
The spectrum of nonperturbative D0-D8brane massless
strings described in detail in our paper 
\cite{flux} resolves this puzzle.

\vskip 0.1in 
In other words, once we work in nine dimensions, enabling the use of
$T_9$ target
space duality transformations and interpolating Wilson line backgrounds, the equivalence 
between the type IB-I$^{\prime}$ and 
heterotic O--E formulations becomes transparent: they are simply alternative 
worldsheet descriptions of identical target spacetime physics. Namely, the 
weak and strong coupling behavior of perturbatively renormalizable 
and anomaly-free theories with sixteen supercharges, and nonabelian
gauge groups $SO(32)$, 
$SO(16)$$\times$$SO(16)$, or $E_8$$\times$$E_8$. It should be emphasized
that the strong-weak coupling duality relating the type IB and heterotic string theories
in ten dimensions, with
$SO(32)$ gauge group, 
holds for the entire spectrum of massive string modes, not merely the massless fields
\cite{flux}. In fact, the massive mode spectrum of the type I$^{\prime}$ ground states with 
D0-D8brane
configurations is best described by developing the isomorphism to the heterotic momentum 
lattice; for details, the reader can consult \cite{flux}. We should note, however, that
this equivalence does {\em not} hold beyond tree-level. As clarified in my recent 
works \cite{ultra}, 
one-loop coupling constant renormalization differs in open and closed string theories
in that both the limit of coincident string vertex operators, as well as the limit of shrinking loop
lengths, can contribute to the ultraviolet divergences of any given string scattering
amplitude. In contrast,
modular invariance ensures the excision of contributions from
the ultraviolet regime to any heterotic closed string scattering amplitude. Of course, since the tree-level 
masses in theories with sixteen supercharges do not receive loop corrections, in this particular
case, 
the distinction becomes a moot point.

\vskip 0.1in 
Are there any additional string backgrounds with sixteen supercharges in dimensions $D$ $\le$ $9$
that are also perturbatively renormalizable? The answer, evident from our discussion in section 2, is 
{\em Yes}, and the original examples
are the supersymmetry preserving CHL orbifolds
of the toroidally compactified heterotic string theories \cite{cp,chl,cl}.
As explained before, the nonabelian 
gauge sector in
the generic CHL orbifold in spacetime dimensions {\em other than four} does not display 
evidence for manifest electric-magnetic
duality. Notice that the special role of four dimensions is also highlighted by consideration of
electric-magnetic duality in the supergravity
sector of either toroidal or CHL compactifications. In generic spacetime dimensions,
the electric two-form potential is 
always present in the perturbative string 
mass spectrum while the dual 
six-form is not. The six-form potential couples to magnetic fivebranes and
it is, therefore, only in four spacetime dimensions that evidence for 
manifest electric-magnetic duality can appear: wrapped fivebranes on six
tori have a pointlike limit, and the conjectured electric-magnetic duality 
takes the form of {\em S-duality} in the low energy N=4
supergravity-Yang-Mills field theory. 

\vskip 0.1in 
In other words, the apparent 
electric-magnetic duality in the gauge sector of the toroidally
compactified heterotic string in generic spacetime dimensions is a red herring: 
since only simply-laced gauge groups appear in the Narain 
moduli spaces, and since in this case the magnetic dual gauge group happens to coincide
with the electric gauge group, the N=4 theories also appeared to be electric-magnetic
self-dual, point-by-point in the moduli space, and in {\em any} spacetime dimension.
The
CHL moduli spaces clarify this issue since they include enhanced symmetry points with 
both simply-laced, and non-simply-laced, gauge symmetry. 
It becomes evident that it is 
{\em only in four spacetime dimensions} that enhanced symmetry points with 
non-simply-laced gauge
groups appear in electric-magnetic dual pairs within the same moduli space, 
an intriguing 
discovery made in our paper \cite{cp}. This particular observation is due to
Joe Polchinski. Notice that at any point in the moduli space, the root lattices of both electric and
magnetic dual groups are accompanied by their respective irreducible highest
weight lattices, a simple 
consequence of modular invariance in every spacetime dimension.
Dual pairs of gauge groups appear within the same moduli space only in four
spacetime dimensions, so that the requisite
spectrum of dual electric and 
magnetic gauge charges can be found at the corresponding enhanced symmetry points, 
in agreement with 4D S-duality.

\vskip 0.1in We should emphasize that the CHL orbifolds describe perturbatively 
renormalizable heterotic string theories that have all of the appealing features of the 10d
$E_8$$\times$$E_8$ and ${\rm Spin}(32)/{\rm Z}_2$ theories. Their generic self-consistency
is especially transparent in the abelian orbifold construction given by David Lowe and
myself in \cite{cl},
generalizing the case of the ${\rm Z}_2$
orbifold described above \cite{cp}. The basic idea is to mod out by the ${\rm Z}_n$
symmetry present at any point in the Narain moduli space where $n$ identical copies of a 
component root-lattice appear as a subspace of the Lorentzian self-dual
lattice, accompanied by a translation in the (10-d,10-d) compactification lattice. For the details,
the reader can consult \cite{cl}. More complicated examples, some of which only invoke effective, 
field theoretic, 
strong-weak dualities can 
be found in \cite{schws}, in addition to the type I and type II abelian CHL orbifolds described in
\cite{flux,land}.
The analog of the abelian 
${\rm Z}_n$ symmetry in type I and type II string backgrounds with Dbranes, 
first pointed out by us in \cite{flux}, is the symmetry under interchange
of $n$ stacks of coincident Dpbranes, with each stack carrying identical worldvolume 
nonabelian gauge group, accompanied by a translation. The result will be a disconnected component
of the type IB moduli space with sixteen supercharges. But it should be noted that,
as in the case of the heterotic string, either vacuum decompactifiies to the ten-dimensional
type I, or type II, string. The examples described in \cite{flux} are the type IB-I$^{\prime}$ strong coupling
duals of the heterotic CHL orbifolds \cite{cp,cl}, further evidence of self-consistency with the 
type IB-I$^{\prime}$/heterotic strong-weak coupling duality conjectures \cite{polwit,witdual,hw}. 

\vskip 0.1in We should emphasize that, as in
the case of toroidal
compactifications, the equivalence of the heterotic and 
type I  supersymmetry-preserving orbifold (CHL) 
compactifications holds at the level of the full string mass spectrum, not only the lowest
lying field theoretic modes. These are two different worldsheet descriptions of identical
target spacetime physics: a perturbatively renormalizable theory of sixteen supercharges
in spacetime dimension less than ten, with an anomaly-free Yang-Mills gauge group 
containing 
$(26-d) -r $ abelian gauge bosons. The integer $r$ can vary from $8$ to a full $26-d$
in four dimensions \cite{chl,cl}. As mentioned above, we should also note that 
the equivalence between the type I and 
heterotic CHL orbifolds holds only at tree-level in the string coupling constant expansion.

\vskip 0.1in How does this picture extend to generic type IIA, type IIB, and M theory 
compactifications preserving sixteen supercharges? Since these theories have 32 supercharges,
we are interested in compactification on spaces with $SU(2)$ holonomy, and the only
nontrivial Calabi-Yau manifold in this class is K3, of complex dimension two. A simpler
possibility enabling precise analysis 
is asymmetric toroidal orbifold compactification, and we will begin our discussion
with this case. Recall that M theory 
compactified on a ${\rm Z}_2$ orbifold gives an anomalous 10d theory \cite{hw}, and the non-anomalous
and, perturbatively renormalizable, extension is nothing but the $E_8$$\times$$E_8$ 
heterotic string in ten dimensions \cite{witdual,hw}. 
In dimensions
$9$ $\le$ $D$ $\le$ $7$, any attempt to break {\em half} of the supersymmetries of the type II
superstrings by asymmetric orbifold compactification {\em without} the introduction of Dbranes,
runs into a clash with modular 
invariance: the modular invariant default is a nonsupersymmetric type 0A or type 0B 
vacuum, which also happens to be tachyonic \cite{polbook}. 

\vskip 0.1in It turns out that Ferrara and 
Kounnas have 
studied an analogous problem in \cite{koun, IIklt}, demonstrating the existence of 
perturbatively renormalizable type II ground states in
four dimensions preserving $N$ $=$ $8$, $6$, $5$, $4$, or $3$ 
supersymmetry in the toroidally compactified type II superstrings on a six-torus. 
In particular, they showed that the 4D $N$$=$$4$ type II modular invariants
preserve either $N$ $=$ $(2_L , 2_R) $, or $N$ $=$ $(4_L ,  0 )$, of the 
$N$ $=$ $(4_L ,4_R )$ supersymmetries 
of the toroidally compactified type II superstrings. How should we interpret these
4D ground states? As regards the supergravity sector alone, the 
former can be identified as a chiral type IB projection of the IIB string theory, projecting to the
symmetric linear combination of left and right moving worldsheet modes. The latter is a heterotic 
chiral projection of the type IIA string, singling out the massless gravitinos from the right moving
world sheet superconformal field theory alone \cite{koun}. Either 
class of $N$ $=$ $4$ ground state
can contain massless Yang-Mills gauge fields, and the number of abelian gauge fields in the low energy
gauge theory is a useful 
hint towards deducing the precise orbifold projections that led to each such ground state. 

\vskip 0.1in It should be emphasized that since care has been taken to 
preserve modular invariance in the type II asymmetric orbifold ground states of  \cite{koun,IIklt}, 
they are likely to describe perturbatively renormalizable 
4d theories with sixteen supercharges, analogous to the heterotic and type I CHL strings 
described above.  A word on nomenclature: the heterotic CHL asymmetric orbifold \lq\lq compactifications" are projections on 
the Hilbert space of the toroidally compactified heterotic string, 
that also {\em preserve} the supersymmetries of the parent string vacuum. The 
\lq\lq asymmetric 
type II orbifold compactifications"  of Ferrara and Kounnas \cite{koun}, and also
Kawai, Lewellen, and Tye \cite{IIklt}, with 4D $N$ $=$ $4$ supersymmetry,  
are a closer, lower-dimensional analog of the 10d chiral projections on the Hilbert spaces of 
the type IIB and type IIA superstrings that give, respectively, the
type IB and heterotic string theories with only half the number of supersymmetries. 
Notice that, while either chiral projection
yields an anomalous $N$ $=$ $1$ theory in ten dimensions, necessitating the addition
of Yang-Mills gauge fields with gauge group $E_8$$\times$$E_8$ or 
${\rm Spin} (32)/{\rm Z}_2$, toroidal compactification 
to four dimensions followed by the identical chiral projections 
directly give self-consistent $N$ $=$ $4$ string ground states.

\vskip 0.1in
The Yang-Mills gauge fields in \cite{koun,IIklt} arise from massless type II closed string 
Kaluza-Klein or closed string winding modes. A further supersymmetry 
preserving (CHL) orbifold of such a
toroidal compactification can lead to new ground states with a Yang-Mills sector containing
fewer abelian gauge fields. The 4D $N$ $=$ $(4_L  ,  0 )$ ground states in the analysis of
\cite{koun,IIklt}
have gauge group $[SO(3)]^6$, $SU(4)$$\times$$SO(3)$, or $SU(3)$$\times$$SO(5)$,
with maximal rank six, precisely as expected for heterotic compactification on a six-torus in
the absence of the ten-dimensional rank 16 gauge group. We emphasize that unlike the 
Narain, or CHL, compactifications \cite{nsw,chl,cp}, these 4D theories\ \lq\lq decompactify" 
self-consistently
to the toroidally compactified type II theory, with twice as many
supersymmetries in six thru ten dimensions, and no Yang-Mills gauge fields. 
The 4D $(2_L , 2_R )$ ground states can have even larger rank gauge groups, since massless 
gauge bosons can arise from both the compactified 
left, and right-moving, conformal field theories. Ref.\ 
\cite{koun} finds 4D $N$ $=$ $4$ ground states with 
gauge groups of rank $22$, $14$, $10$, $6$, and $2$. The latter four cases are likely to
correspond to supersymmetry preserving (CHL) orbifolds of the Hilbert space of the
ground state with $22$ abelian gauge fields. In 6D, the analogous maximal rank gauge group
obtained in \cite{koun} was rank 20. The scalars of the maximal rank $N$ $=$ $4$ ground state
in \cite{koun} 
parameterize the manifold $[SU(1,1)/U(1)]$$\times$$[SO(6,22)/SO(6)\times SO(22)]$, precisely 
as in 4D toroidal compactifications of the $SO(32)$ type IB-heterotic string theory. 

\vskip 0.1in It would be
nice to have a systematic classification of the asymmetric toroidal orbifolds of the type IIA and
type IIB superstring theories 
in every dimension $2$ $\le$ $D$ $\le$ $9$, for theories with 32, 16, or 12 supercharges;
the further reduction to 8 or fewer supercharges leads to a well-known, rapid proliferation of 
solutions. 
For example,
based on our discussion here, it is apparent
that enhanced symmetry points with non-simply laced gauge symmetry must 
exist in the type II moduli spaces. It would be nice to have a classification of the enhanced gauge 
symmetry points in each dimension, and to verify the appearance of S-duality in every 4D  
$N$ $=$ $4$ case. To reiterate the general theme of this section, we conclude that
the perturbatively
renormalizable type I, type II, or  
heterotic, string ground states with sixteen supersymmetries, in nine dimensions and
below, are simply alternative
worldsheet conformal field theory descriptions of identical target spacetime physics.

\vskip 0.1in To drive home this point, it is helpful to consider what additional insight
might be gained from the study of 
type II string compactification on smooth K3 surfaces? $SU(2)$ holonomy implies that we
have a theory with sixteen supersymmetries, and the massless scalars of the non-chiral 
 $N$ $=$ $2$ supergravity in six dimensions parameterize a
 space that is locally equivalent to $[SO(20,4)/SO(20)\times SO(4)]$ \cite{seiberg}. 
 The gauge group at generic points in the moduli space is simply $[U(1)]^{24}$. 
 The 6D $(1_L , 1_R )$ orbifold compactifications of \cite{koun} with nonabelian gauge symmetry 
 correspond to special points in the moduli space
 of {\em quantum} K3 surfaces where, upon tuning the (3,19)-components of the so-called
 \lq\lq B" field to particular values, the type IIA string can acquire a variety of 
 enhanced gauge symmetries \cite{k3,aspdual}: the corresponding classical K3 surface 
 has an orbifold quotient singularity that falls within an A-D-E classification, and the role of
 the B-field is crucial in its quantum resolution. For details, the reader can consult 
 the papers by Aspinwall and Morrison 
 \cite{k3,aspdual}, which include an exposition of the remarkable fact that 
 even the {\em global} structure of the moduli space of the type II string compactified 
 on K3 agrees precisely with that of the heterotic string compactified on a four-torus. 
 
\vskip 0.1in 
These observations underlay the extensive exploration of a heterotic-type IIA string-string
strong-weak effective, field theoretic, duality in six dimensions 
\cite{sen1,schws,cl,aspdual}. The
detailed encoding of the precise geometric data required to specify a quantum K3
surface is beautifully captured by the (4,20)-dimensional Lorentzian self-dual lattice 
that also specifies the background fields of a heterotic toroidal compactification \cite{nsw,k3,aspdual}.
This isomorphism enabled us \cite{cl} to identify the precise
supersymmetry preserving CHL orbifold action on the cohomology lattice 
that gives a new moduli space of quantum K3 surfaces of smaller dimension. A quantized
background one-form potential in the IIA Ramond-Ramond sector plays a crucial role 
in establishing this equivalence \cite{sen1,schws,cl}. The
classification of such symplectic automorphisms of the cohomology lattice of K3 
surfaces due to Nikulin \cite{nik}, in particular, provides a detailed 
enumeration of the perturbatively renormalizable 4D $N$ $=$ $4$ 
type I-type II-heterotic ground states
that follow as orbifold projections of the ten-dimensional $N$ $=$ $1$ superstrings.

\vskip 0.1in 
As a final comment, it would be interesting to complete the derivation 
of the modular invariant one-loop
vacuum amplitude for the CHL orbifolds in the lattice representation, 
with lattice momenta parameterized by background field vevs \cite{nsw}.
In both our works 
\cite{chl,cp,cl}, and in the Ferrara-Kounnas \cite{koun} analysis, the 
constraints from one-loop 
modular invariance have indeed been 
verified in both fermionic and orbifold formalisms. 
But it is an explicit lattice representation analogous to \cite{nsw} that would provide the
clearest physical insight because of the background field parameterization. Such a
representation would also shed light on the isomorphism to quantum K3 geometry
when such an isomorphism is available, see the related discussion in \cite{aspdual}.

\section{Sharpening the Boundaries of String Theory}

\vskip 0.1in 
We will end by presenting two mathematical curiosities easily deduced from 
the fermionic current 
algebra representation, but whose physical significance remains a puzzle.
We emphasize that we do not as yet have an understanding of these solutions as 
either geometrical, or even non-geometrical, CHL orbifolds. The existence
of such stray exactly solvable conformal field theory solutions is therefore a 
nuisance, but surely also an opportunity, to learn about the boundaries of string 
consistency. It is in this spirit that we have decided to include them in this paper.
Let us address
a recent question put forward by Martin Rocek: {\em Are
there any 6D N=2 or 4D N=4 heterotic string vacua lacking 
the familiar right-moving
abelian gauge fields present in all toroidal compactifications: four in 6D,
six in 4D, (and eight in 2D, with an N=8 supersymmetry)?} 

\vskip 0.1in As explained in the introductory sections, since we lack both a 
complete classification of (22,6)-dimensional Lorentzian self-dual lattices, 
as well as a classification of their supersymmetry preserving automorphisms, it is 
very difficult to
formuate a no-go proof of such a conjecture. However, it is sometimes possible
to construct an explicit counter-example. It turns out that the fermionic
CHL strings described in my paper with Lykken and Hockney 
\cite{chl} included a 4D N=4 example 
which lacked all of the 22 left-moving abelian gauge fields expected in the generic
toroidal compactification \cite{chl}. We remind the reader that such an 
N=4 theory has no massless scalar fields other than the dilaton from the 
gravity multiplet: {\em the moduli space of massless scalars is trivial at weak 
string coupling, and we apparently have an isolated 4D N=4 heterotic string vacuum}. 
Needless to say, this solution does not fall within the class of CHL orbifolds that
can decompactify to the ten-dimensional heterotic string.\footnote{Evidence for 
non-geometric symplectic orbifolds which may not decompactify to ten dimensions 
was also given by us in \cite{clm}, but that construction appears a bit contrived.}
The question is whether one can also 
eliminate all six right-moving abelian gauge fields. Such exotic perturbatively
renormalizable and
ultraviolet finite string solutions would be extremely difficult to guess at
by purely field-theoretic supergravity   
considerations. The self-consistency of such a cft solution rests, instead, on 
the existence of an exactly solvable description of an all-orders in 
$\alpha^{\prime}$ 
background of heterotic string theory given by the fermionic current algebra
representation \cite{chl}. 

\vskip 0.1in 
As an example of the utility of such stray exactly solvable 
fermionic current algebra solutions, we will 
point out that one can deduce an immediate answer 
confirming the 2D 
N=8 conjecture, but without ruling out either the 4D N=4, or 6D N=2, conjectures. 
However, as will become clear below, it appears very unlikely that solutions for
the latter two conjectures 
can exist, and 
the reason
is as follows. Consistency with modular invariance, and the existence of unambiguous
fusion rules for the twisted current algebra, requires, as
was shown in \cite{consist}, that the worldsheet superconformal field theory 
necessarily contain sectors with 
specified blocks of self-consistent boundary conditions.
The ${\rm Z}_2$
twisted worldsheet Majorana fermions, each with central charge $\half$, 
are required to appear in blocks of 16, 24, 28, 32, 36, 
40, 44, 48... this exhaustive enumeration could be continued to arbitrarily
large-sized fermion blocks, at the cost of additional computer time. 

\vskip 0.1in For example, the block of 16 appears among the left-movers in all of the 
fermionic current algebra solutions describing points in the moduli space of the 
Chaudhuri-Polchinski ${\rm Z}_2$ orbifold \cite{cp}, giving a reduction of eight left-moving
abelian gauge fields in the massless spectrum. The novel 4D N=4 solution described
above utilizes, instead, the block of 44 left-moving 
twisted Majorana fermions \cite{chl}, eliminating 
all 22
left-moving abelian gauge fields. What are the additional restrictions on the boundary
conditions on right-moving
Majorana fermions as a consequence of the triplet constraint from 
worldsheet supersymmetry? In four dimensions, and in light-cone gauge, 
the counting of worldsheet degrees of freedom goes as follows.
Recall that the right-moving superconformal field
theory has central charge 1+9, where we have singled out the c=1 unit that carries
the two (transverse) spacetime charges of the gravitinos in 4D. 
The internal cft with total central charge c=9 arises from fermionizing  
six right-moving chiral bosons, namely, from the six-torus, in addition to their six
right-moving worldsheet Majorana fermion superpartners. In all, we have a total of 18
internal, right-moving Majorana fermions. In heterotic conformal field theory
solutions with 16 conserved spacetime supercharges, the 
{\em triplet constraint} of \cite{klst,consist} is nothing but
the requirement that the gravitinos live in the spinor of an SO(8): this removes a
c=3 unit from the internal c=9 fermionic cft, combining it with the c=1 unit to give the root
and irreducible weight lattices of SO(8) at Kac-Moody level one, namely, spinor, 
conjugate spinor, 
and vector. The remnant block of 12 Majorana fermions
with central charge six cannot meet the modular invariance restrictions referred
to above, since there are no self-consistent ${\rm Z}_2$-twisted fermion blocks of 12.
It is evident that the 6D N=2 case is even more constrained, since the remnant 
right-moving block has only 8 Majorana fermions! Given that the ${\rm Z}_2$ orbifold, with 
its minimum reduction of eight abelian gauge fields, is the simplest known 
CHL orbifold, we think it unlikely that the generic orbifold analysis can remove this
basic hurdle from modular invariance. However, our argument does not in itself constitute a 
definitive no-go proof.

\vskip 0.1in Upon compactifying further to 2D, the 
restrictions from modular invariance loosen up, and appear to allow a 
viable solution for the fermionic current algebra.
In two dimensions, the counting of worldsheet degrees of freedom is
as follows: in light cone gauge, all right-moving Majorana fermions are
now internal, and we have a total of 24 fermions since we must fermionize two
additional chiral bosons. Reserving
a c=4 unit from the internal conformal field theory
for the gravitino embedding as before, we are left with a block
of 16 right-moving internal Majorana fermions. We can readily satisfy
the triplet constraint simultaneous with the requirements from modular
invariance: choosing a block of 16 right-moving twisted Majorana fermions,
together with the block of 48 left-moving twisted Majorana fermions described
in \cite{consist}, appears to 
give a 2D N=8 heterotic string vacuum with no abelian gauge
fields originating in either left-,  or right-moving, worldsheet conformal field
theories. Whether this fermionic solution can be given an orbifold interpretation remains
to be seen, apart from what that might imply for our generic observations on 
decompactification.

\section{Conclusions}

\vskip 0.1in Understanding 
the symmetry principles, and the fundamental
degrees of freedom in terms of which nonperturbative String/M theory 
can be formulated, is an important focus of ongoing research in
theoretical high energy physics. Elucidating the web of heterotic-type I-type II
CHL compactifications preserving sixteen or fewer supercharges 
can play a significant role in guiding such work  
since it determines precise boundaries for what we mean by 
{\em string consistency}.

\vskip 0.1in A repeated theme in section 3 of this paper was the unity of the different
superstring theories: alternative worldsheet descriptions, type I, type II, or heterotic,
of identical target spacetime physics, although one must take care to note that
the type I/heterotic equivalence holds only at tree-level in the string coupling constant. 
Nevertheless, since the
mass spectrum in theories with sixteen supercharges receives no loop corrections,
the equivalence remains a powerful conclusion.
The web of CHL orbifolds described in this paper reiterates this point, supplementing
the known heterotic examples 
with both type I, and type II, perturbatively renormalizable
backgrounds preserving sixteen supercharges \cite{flux}, also
making contact with previous results of 
Ferrara and Kounnas \cite{koun}. These results go a long way towards establishing
self-constency with the type I-heterotic-M strong-weak coupling duality conjectures,
in a multitude of disconnected moduli spaces, and in diverse spacetime dimensions. 

\vskip 0.1in We strongly believe that a more systematic investigation
will only confirm that the generic 
4D type II moduli spaces with
$N$ $\le$ $4$ are precisely isomorphic to the CHL toroidal orbifolds of the 
heterotic/type I string theories described in sections 2 and 3. We emphasize that
this conclusion is self-consistent  
with the strong-weak coupling duality conjectures \cite{witdual,polwit,hw}.
Recall that the type IIA and type IIB string theories have 
Ramond-Ramond sectors with pform gauge potentials in the supergravity sector,  
apparently an alternative route to orbifolding
for the construction of type II ground states with sixteen
supercharges. We have found, however, that requiring the absence of Ramond-Ramond
sector tadpoles leaves the possibilities for new solutions severely constrained. 
The introduction of a Dpbrane in any toroidal
type II compactification breaks exactly half the supersymmetries, for any $p$, and  
consistency requires that we introduce 32 of them in order for the cancellation of 
Ramond-Ramond sector tadpoles \cite{polbook}. In nine dimensions, 
this gives only the standard $O(32)$ type IB, and T-dual type IA, backgrounds
with, respectively, 32 D9branes or 32 D8branes. 
An intersecting brane 
configuration of (p,p-8) branes preserves all of the supersymmetries so, as explained
in section 3, we can also construct a type I$^{\prime}$ background with D0-D8branes
and $E_8$$\times$$E_8$ 
gauge group \cite{flux}. Finally, a sequence of $p$ T-dualities can map the background 
with 32 D9branes 
to flat space backgrounds with $p$ transverse bulk coordinates, and $10-p$ longitudinal
worldvolume coordinates, and we have a perturbatively renormalizable ground state
with 16 supersymmetries and $O(32)$ D(9-p)brane worldvolume gauge fields. The 
introduction of gauge symmetry breaking Wilson 
lines, and quantized background antisymmetric two-form potential 
\cite{flux}, enables supersymmetry preserving CHL orbifold 
compactification, giving the generic type I or type I$^{\prime}$ ground state with sixteen
supersymmetries. 

\vskip 0.1in A comprehensive classification of CHL compactifications
in four dimensions with N = 8, 4, 3, 2, 1, and (zero) remains an important goal,
although it appears unlikely to materialize in the near future as explained
earlier. Some questions of interest that may be answerable even without a full
classification include the following.
Vafa: 
{\em Are there any 4D N=2 superstring vacua without any additional massless 
scalar fields other than the dilaton?} Very likely, yes. But it has not been
systematically explored.
Vafa, Rocek, Niewenhuizen, Chaudhuri: {\em 
What is known about
the 4D N=3 theories coupled to matter? Do all of these theories contain points
in the moduli space with an extended N=4 supersymmetry? Does this always 
require decompactification (a degeneration) of one, or more, cycles of the torus?} 
We believe this last issue would be extremely interesting to
address, and we will leave it for future work. Asymmetric 
orbifolds of the type II 
superstrings with 4D N=3 supersymmetry appear among the Ferrara-Kounnas
examples
described in section 3 \cite{koun}, and include examples of identifiable CHL orbifolds. 

\vskip 0.2in \noindent {\bf Acknowledgments}: I would like to thank Joe Polchinski, 
Joe Lykken, David Lowe, and also George Hockney, for their collaboration
on the early research described here. I would also like to acknowledge useful
discussions
with Paul Aspinwall, Costas Kounnas, and Andrew Frey. 
It is a pleasure to thank the organizers and 
participants of the {\em 3rd Simons
Workshop in Mathematics \& Physics}, especially Cumrum Vafa and Martin Rocek,
for a stimulating meeting.

\end{document}